\documentclass{article}
\usepackage[T1]{fontenc}
\usepackage[utf8]{inputenc}
\usepackage{multirow}
\usepackage{ismir}
\usepackage{amsmath,cite,url}
\usepackage{graphicx}
\usepackage{color}
\usepackage{xcolor}
\usepackage{booktabs}
\usepackage{float}
\usepackage[bookmarks=false]{hyperref}
\usepackage{amssymb}

\title{MambaVoice: Lightweight Audiovisual Singing Voice Separation via a Hybrid Mamba-Transformer Model}

\multauthor
{Adithi Shankar$^{1}$ \hspace{1cm} Gopika Krishnan$^{1}$ \hspace{1cm} Gloria Haro$^{2, 3}$} { \bfseries{ Xavier Serra$^{1}$ \hspace{1cm} Martín Rocamora$^{1}$ \hspace{1cm}}\\
  $^{1}$ Music Technology Group, Universitat Pompeu Fabra, Barcelona, Spain \\
 $^{2}$ Intelligent Multimodal Vision Analysis Group, Universitat Pompeu Fabra, Barcelona, Spain \\
 $^{3}$ Serra Húnter Fellow Programme, Universitat Pompeu Fabra, Barcelona, Spain\\
}

\def\authorname{A. Shankar, G. Krishnan, G. Haro, X. Serra and M. Rocamora}

\begin{document}

\maketitle

\begin{abstract}
\noindent
Isolating a target singing voice from a music video remains challenging, particularly in the presence of multiple vocalists and dense instrumental accompaniment. We propose MambaVoice, a lightweight audiovisual framework that leverages a hybrid Mamba–Transformer architecture for targeted singing voice separation. The model jointly encodes audio and visual streams using an attention-based band-split audio encoder and a spatio-temporal graph convolutional network (ST-GCN) for facial motion features. These modalities are fused through a multiplicative gating mechanism, enabling visual cues to selectively modulate audio representations. The fused features are processed by a hybrid backbone that combines Transformer self-attention with Selective State Space Models (SSMs), achieving efficient long-range temporal modeling with linear complexity. We evaluated MambaVoice on the Acappella and URSing datasets under challenging conditions, including mixtures with interfering singers. At 16.2 million parameters, the model demonstrates comparable performance, achieving 14.18 dB SDR on Acappella and strong cross-dataset performance on URSing, comparable to larger models at a fraction of the parameter count. These findings highlight the effectiveness of hybrid SSM–attention architectures for scalable, efficient audiovisual source separation, suggesting they are well-suited as lightweight components within larger pipelines. We conduct a perceptual study that further supports our improvements in objective metrics. We provide our implementation online.\footnote{Code available at: \url{https://github.com/adithishankar19/mamba_voice.git}}
\end{abstract}

\section{Introduction}
\label{sec:intro}

The task of Audiovisual Singing Voice Separation (AV-SVS) involves isolating a target singing voice from a mixture containing accompanying instruments and backing vocals, guided by the visual cues of the target performer. Unlike speech separation, which often deals with intermittent noise, singing voice isolation requires disentangling the target from the interference of musical instruments and backing vocals. A significant challenge in this domain is target swapping, in which a model inadvertently tracks an interfering singer because of timbral and spectral similarities. This is particularly prevalent in professional studio recordings, where backing vocals are mixed with the lead, creating dense overlapping harmonics that audio-only models struggle to separate. 

Early state-of-the-art methods in audiovisual separation, such as  \cite{ephrat2018looking}, established the effectiveness of target-guided separation in speech. These models utilize video frames associated with a specific face as an anchor for the vocal signal.  However, singing exhibits a significantly more intricate temporal and spectral structure than speech, primarily due to the sustained melodic contours and the overlap with instrumental accompaniment. State-of-the-art methods in AV-SVS \cite{montesinos2021cappella,montesinos2022vovit} have heavily relied on Deep Convolutional U-Net based networks \cite{ronneberger2015u} or Transformer based architectures \cite{vaswani2017attention} to jointly learn audio and video modalities for separation. While effective, these models often suffer from distinct shortcomings. Although U-Net based architectures are significantly lighter in terms of computational complexity, they are constrained by a fixed and often insufficient receptive field, which hinders their ability to model the long-range temporal dependencies inherent in musical structures. Conversely, while Transformers provide the necessary global context, they incur a quadratic computational cost and are significantly heavier during training. 

In this work, we propose a novel audiovisual separation system based on a Hybrid Mamba-Transformer backbone, designed to tackle the quadratic memory bottleneck of standard Transformers. By stacking Selective State Space Models (Mamba blocks) \cite{gu2024mamba} for linear-time temporal modeling with Transformer blocks for global spectral fusion, our architecture allows for the processing of long-form musical sequences without affecting global context. To achieve computational efficiency, the proposed system utilizes an attention-based Band-Split module for audio encoding, which partitions the mixture into sub-band features corresponding to harmonic regions. A Spatial-Temporal Graph Convolutional Network (ST-GCN) \cite{yan2018spatial} models the dynamics of facial keypoints to extract motion features relevant to the singing performance. These multimodal representations are integrated via a multiplicative gating mechanism, in which the visual features modulate the audio embeddings to identify relevant features that guide the separation. 

At 16.2M parameters, the model achieves an SDR of 14.18 dB on the unseen-unheard testing set of the Acappella dataset \cite{montesinos2021cappella}, demonstrating that the Hybrid Mamba-Transformer architecture provides effective separation without the memory costs of pure attention-based systems. 

Beyond the specific model, our ablations point to a design principle that should transfer to other audiovisual separation architectures. At fixed depth and parameter budget, placing the linear-time recurrence before attention (M-T) consistently outperforms the reverse arrangement (T-M) across both datasets and both interference regimes. Since the two variants differ only in the ordering of the blocks, this indicates that selective SSM layers are most useful as a temporal front-end that stabilises and compresses the sequence for a subsequent global spectral mixer, rather than as a refinement of already-attended features.

\section{Related Work}
\label{sec:related work}

Singing voice separation based on audio has evolved from signal processing techniques like Non-Negative Matrix Factorization \cite{rohlfing2016nmf,ewert2012using} to Deep Learning (DL) approaches. Initial DL benchmarks were established by models such as Open-Unmix \cite{stoter2019open} and Spleeter \cite{hennequin2020spleeter}, which utilized Bi-LSTMs and U-Nets to estimate time-frequency masks. Subsequent research shifted toward time-domain modeling to better preserve phase information, exemplified by Demucs \cite{defossez2019demucs}, which is a U-Net-based architecture with a Transformer at the bottleneck, DPTNet \cite{woo2021speech} utilized dual-path transformers to capture long-range temporal dependencies within the audio signal. Despite these advancements, audio-only systems are fundamentally limited by spectral overlaps. When lead and backing vocals occupy similar frequency ranges with identical timbres, these models lack sufficient discriminative context, frequently resulting in target swapping or artifacts. While modern frameworks like BandSplitRNN \cite{luo2023music} have improved feature extraction by partitioning the Short-time Fourier Transform (STFT) into harmonically-informed sub-bands, their reliance on recurrent or attention-based backbones creates a computational bottleneck as sequence lengths increase.

A parallel branch of research has addressed multi-singer scenarios through Permutation Invariant Training (PIT). PIT-based models \cite{sarkar2021vocal, petermann2020deep} are often applied to choral or ensemble separation, allowing a network to isolate multiple sections of choral music like soprano, alto, tenor and bass. However, while PIT is effective for multiple sections in a choir, it does not inherently solve the identity problem. In professional recordings, a clear distinction must be maintained between a lead vocalist and a backing choir. PIT-based systems cannot reliably ensure that the lead vocal remains in a consistent output channel throughout a track, particularly during periods of overlap or silence, because the model lacks an external reference to anchor the target singer's identity.

Audiovisual separation addresses this identity ambiguity by incorporating visual motion as a grounding signal for the target source. Using visual cues, these models can mitigate the target-swapping issues inherent in PIT and audio-only frameworks. This approach has been facilitated by the release of specialized multi-modal corpora, such as the URSing dataset \cite{li2021audiovisual}, which established the initial correlations between vision and vocal acoustics, and the Acappella dataset \cite{montesinos2021cappella}, which provides a large-scale collection of solo singing performances across diverse genres. These datasets enabled the development of architectures such as Y-Net \cite{montesinos2021cappella}, which fused spectrograms and facial gesture in a shared latent space, and VoViT \cite{montesinos2022vovit}, which utilized cross-modal transformers that use attention-based mechanisms to attend to video features relevant to audio to guide separation. While these attention-based models achieve high accuracy, the primary limitation of these Transformer-based approaches is their quadratic computational complexity. 

As an alternative to the computationally expensive Transformers, Selective State Space Models (SSMs) with linear complexity, such as Mamba \cite{gu2024mamba}, have emerged as a viable solution for long-sequence modeling. By utilizing a hardware-aware selective scan mechanism, Mamba facilitates the modeling of extensive temporal dependencies with $O(L)$ complexity, effectively bypassing the memory constraints associated with standard self-attention. Although Mamba excels at capturing long-range temporal dependencies, it may lack the dense, global spectral modeling capabilities that Transformers provide for complex musical mixtures. This limitation has led to the emergence of hybrid architectures in other domains such as Jamba \cite{lieber2024jambahybridtransformermambalanguage} in large language modeling and MambaVision \cite{hatamizadeh2025mambavision} in image analysis which interleave Mamba blocks with attention layers to balance hardware-efficient recurrence with global context. However, the application of such hybrid SSM-Transformer frameworks remains relatively unexplored in the context of audiovisual source separation. This work bridges that gap by proposing a hybrid backbone that leverages the efficiency of selective scans for temporal tracking while retaining attention mechanisms for high-resolution spectral fusion, resulting in a system that is computationally lightweight while remaining competitive with  larger Transformer-only baselines.

We note that a separate line of work reduces attention's cost directly with efficient transformers, with Linformer \cite{wang2020linformerselfattentionlinearcomplexity} and Performer \cite{choromanski2022rethinkingattentionperformers} using low-rank or kernel approximations and Reformer \cite{kitaev2020reformerefficienttransformer} restricting attention to sparse, hashed patterns. We opt for a hybrid design instead, which retains full attention where dense global spectral modeling is needed rather than approximating it throughout.

\begin{figure*}[t] 
    \centering
    \includegraphics[width=\textwidth]{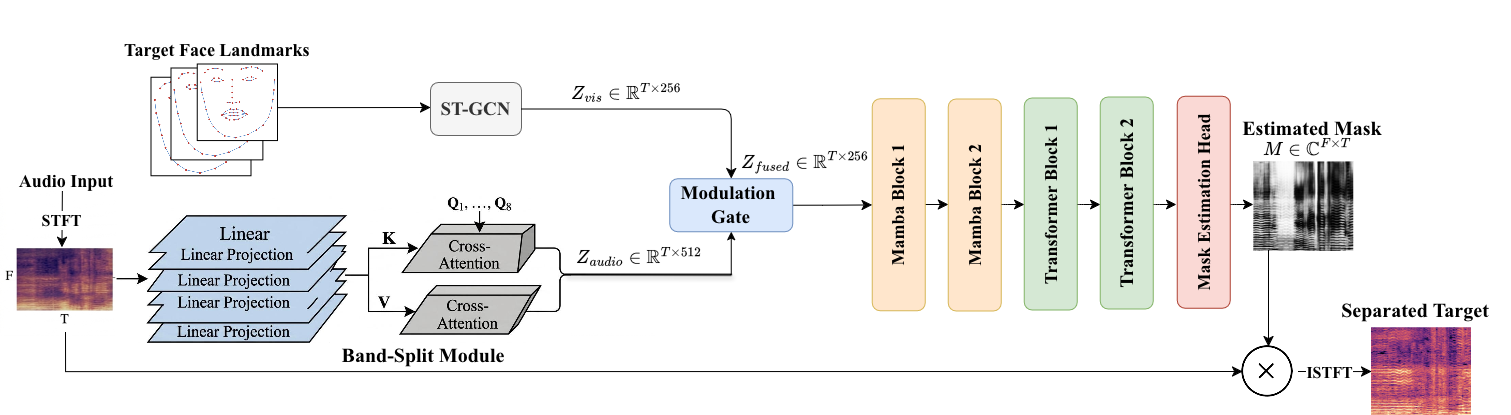}
    \caption{Architecture of the proposed Mamba-Transformer (M-T) hybrid model. The system fuses target facial landmarks with acoustic features to estimate a complex ratio mask for vocal separation.}
    \label{fig:pipeline}
\end{figure*}

\section{Methodology}
\label{sec:methodology}

Figure \ref{fig:pipeline} illustrates our proposed architecture, with individual components detailed in the following sections.

\subsection{Attention-Based Band-Split Audio Network}
Our pipeline transforms the input waveform into a complex spectrogram $S \in \mathbb{C}^{F \times T}$ with $F=256$ frequency bins. CNN-based encoders such as Spec2Vec \cite{ephrat2018looking,montesinos2022vovit} apply identical filters across the entire frequency axis, which is computationally costly and weights all spectral components equally regardless of musical significance. Instead, we propose an Attention-Based Band-Split Module that exploits the harmonic structure of singing voices via logarithmic frequency scaling. The spectrum is partitioned into $K=8$ non-uniform sub-bands: (0,4), (4,12), (12,28), (28,52), (52,84), (84,128), (128,192), (192,256), giving finer frequency resolution in the lower bins (0--28), where $f_0$ and the primary harmonics concentrate, and wider bands above (28--256) to favor temporal resolution.

Each sub-band passes through a 1D convolution (kernel size 3), SiLU activation, and adaptive average pooling, then a linear projection and normalization into a $D=128$ latent space, compressing 256 bins into 8 band embeddings. A Global Band Interaction layer adds learnable positional embeddings and applies Multi-Head Self-Attention across bands. Unlike standard BandSplit designs that treat bands in isolation, this models long-range cross-band correlations, e.g. between the fundamental pitch and distant higher-order harmonics. The resulting context-aware embeddings are concatenated into a dense, efficient input for the Mamba-Hybrid backbone.

\subsection{Spatio-temporal Graph Convolutional Network}
To extract facial motion features, we use the same Spatio-Temporal Graph Convolutional Network (ST-GCN) and 68 2-Dimensional facial landmarks as in \cite{montesinos2021cappella}. The network models the dynamics of facial gestures to produce a latent identity anchor, which serves as a behavioral signature to guide the separation process.




\subsection{Audiovisual Fusion via  Multiplicative Gating}

To resolve source ambiguity in mixtures, we employ a multiplicative gating mechanism that facilitates a dynamic, top-down modulation of the audio stream. Unlike standard additive fusion, which passively concatenates multimodal features, this approach utilizes the visual signal as a learned mask to selectively weight the audio stream. The fusion process begins by aligning the visual embedding $Z_{vis}$ from the ST-GCN with the audio representation $Z_{audio} \in \mathbb{R}^{T \times 512}$, where $T$ denotes the number of time frames. Since these modalities have different dimensionalities, $Z_{vis}$ is first projected via a linear layer to match the audio dimensionality. To transform the visual features into a soft-decision signal, $Z_{vis}$ is projected using a learnable weight matrix $W_g$ and a bias vector $b_g$ into a soft-decision signal to produce a gating mask $G \in [0, 1]^{T \times 512}$:\begin{equation}G = \sigma(W_g Z_{vis} + b_g).\end{equation}The final fused representation $Z_{fused}$ is then computed through the element-wise Hadamard product:\begin{equation}Z_{fused} = Z_{audio} \odot G.\end{equation}By treating the motion embeddings as a weighting function, the model ensures that the separation backbone is conditioned on audio segments that exhibit a relevant gestural correlation with the target, effectively removing interferences of the background vocalists and unrelated acoustic sources or noise.







\subsection{Mamba-Hybrid Backbone}

\begin{figure}[h]
    \centering
    \includegraphics[width=0.5\columnwidth]{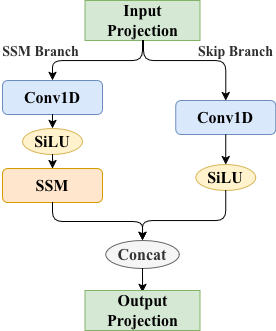}
\caption{Mamba block. The SSM branch applies depthwise conv and SiLU followed by selective scan; the skip branch bypasses the scan with depthwise conv and SiLU only. Outputs are concatenated and projected.}
\label{fig:mamba_mixer}
\end{figure}

Following MambaVision~\cite{hatamizadeh2025mambavision}, each SSM block linearly projects the input and splits it into an SSM branch and a symmetric path, as illustrated in Figure \ref{fig:mamba_mixer}, both using non-causal depthwise convolutions.

The SSM branch applies a depthwise convolution with a SiLU activation  followed by a selective scan~\cite{gu2024mamba}. Let $\mathbf{x}_t \in \mathbb{R}^{512}$ denote the input to the SSM at time frame $t \in \{1, \ldots, T\}$. This scan propagates a hidden state $\mathbf{h}_t$ across frames as a linear dynamical system:
\begin{equation}
    \mathbf{h}_t = \bar{\mathbf{A}}_t \mathbf{h}_{t-1} + \bar{\mathbf{B}}_t \mathbf{x}_t, \qquad \mathbf{y}_t = \mathbf{C}_t \mathbf{h}_t + \mathbf{D}\mathbf{x}_t
\end{equation}
where $\mathbf{C}_t$ is the output projection and $\mathbf{D}$ represents a skip connection. The discrete matrices  $\bar{\mathbf{A}}_t$ and $\bar{\mathbf{B}}_t$ are discretizations of the continuous-time state transition matrix $\mathbf{A}$ and  input projection $\mathbf{B}$. This discretization is controlled by  a learned per-frame step size $\Delta_t$, 
making the scan selective, that is:
\begin{equation}
\begin{aligned}
    \bar{\mathbf{A}}_t &= \exp(\Delta_t \mathbf{A}) \\
    \bar{\mathbf{B}}_t &= (\Delta_t \mathbf{A})^{-1} (\exp(\Delta_t \mathbf{A}) - \mathbf{I}) \cdot \Delta_t \mathbf{B}
\end{aligned}
\end{equation}

The symmetric path applies a non-causal depthwise convolution with SiLU only, serving as a residual feature path that bypasses the sequential scan. Both outputs are concatenated and processed through a feed-forward network before being passed to the Transformer blocks for global spectral refinement.



\subsection{Mask Estimation and Signal Reconstruction}

The refined latent representation $Z$, produced by the hybrid backbone, is passed to a dedicated complex separation head. This head utilizes a $1\times1$ convolutional layer to estimate a complex-valued ratio mask ${M}={M}_{R}+i{M}_{I}$. The target complex spectrogram $\hat{S}_{target}$ is subsequently recovered through complex element-wise multiplication between the estimated mask and the original mixture spectrogram $X_{mix}$, such that $\hat{S}_{target}={M}\cdot X_{mix}$. Finally, the time-domain singing voice is reconstructed by applying the Inverse Short-Time Fourier Transform (ISTFT).

\section{Experiments}
\label{sec:experiments}

\subsection{Training Datasets and Data Augmentation}

The model is primarily trained on the Acappella dataset, which provides isolated vocal tracks with synchronized video. We extract 68-point 2D facial landmarks from each frame using Alphapose \cite{fang2022alphaposewholebodyregionalmultiperson}, a lightweight pose estimator that keeps the visual front-end inexpensive. To simulate real-world musical environments, we implement a data augmentation pipeline in which target vocals are mixed with instrumental stems from the MUSDB18 \cite{rafii2017musdb18} training set and a separate vocal-like training set from the Audioset dataset \cite{gemmeke2017audio}.
\subsection{Testing Dataset}
We test on the unseen-unheard testing set from the Acappella dataset with random accompaniment mixing from the testing set of the MUSDB18 dataset and vocal-like sounds from the Audioset dataset  that has not been part of the training set. 
To evaluate cross-dataset generalization, we utilize the URSing dataset \cite{li2021audiovisual}, which contains high-quality isolated vocal tracks synchronized with upper-body video of the soloists. A unique challenge of URSing is the presence of backing vocals within the accompaniment tracks, which creates high spectral overlap and source ambiguity. This dataset serves as a rigorous test, as it requires the model to leverage visual cues to isolate the primary soloist from both instruments and competing vocal textures in a cross-domain scenario. Across all experiments, $N$ denotes the total number of data samples evaluated.


\subsection{Training Protocol and Baselines}
We distinguish between the interference ratio used during training and the interference ratio used at evaluation. Prior audiovisual singing voice separation systems such as Acappella~\cite{montesinos2021cappella} and VoViT~\cite{montesinos2022vovit} typically train with an interfering voice in 50\% of the mixtures. We use this reported setting for the main VoViT baseline. During development, we found that training the proposed M-T-Hybrid model with an interfering singer in every mixture improved robustness to continuous vocal overlap. Therefore, our main M-T-Hybrid model is trained with a 100\% interference curriculum. To make this comparison explicit, we also retrain VoViT under the same 100\% interference curriculum, denoted VoViT 100\% in the results.

At evaluation time, we report two complementary regimes. Table~\ref{tab:results_50} evaluates models under the standard 50\% interference condition, where the interfering singer is present only in part of the test mixtures. Table~\ref{tab:results_100} reports a 100\% interference stress test, where a competing vocal source is present in every test mixture. Apart from the interference curriculum, all models use the same 16 kHz sampling rate, STFT parameters, data splits, and augmentation pipeline based on Acappella, MUSDB18, and AudioSet.

\subsection{Architectural Ablations}
All variants use a fixed depth of four layers for a controlled comparison of architectural efficiency. Pure Mamba (four selective scan blocks) tests a purely recurrent state-space backbone, while Pure Transformer (four self-attention layers) provides a global spectral modeling baseline without linear-time recurrence. To probe the effect of structural ordering, we compare our proposed Mamba-Transformer (M-T), two Mamba blocks followed by two Transformer layers to prioritize early temporal distillation, against the reversed Transformer-Mamba (T-M). We further replace our band-split encoder with Spec2Vec, which applies uniform convolutional filters across frequency, and evaluate an Audio-only variant without the ST-GCN stream to isolate the gains from multimodal grounding in resolving acoustic ambiguity.

\subsection{Training  Objectives and Configurations}
The model is optimized using a weighted combination of Time-domain $L_1$ loss and Multi-resolution STFT loss, ensuring high-fidelity reconstruction of both the transient envelopes and the harmonic fine-structure. The training is executed on a single NVIDIA RTX 4090 GPU with a batch size of 4. We utilize the AdamW optimizer ($lr=2\times10^{-4}$) with a Linear Warmup for the first 5 epochs to stabilize the ST-GCN and Mamba-Hybrid layers. This is followed by a Cosine Annealing decay. To prevent overfitting and manage convergence on the complex 100\% interference task, we employ ReduceLROnPlateau, which reduces the learning rate by 50\% if the validation Source-to-Distortion Ratio (SDR) fails to improve for 10 consecutive epochs. All audio is resampled to 16 kHz, and spectrograms are computed using an STFT window size of 512 with a hop length of 160.

\section{Results}
\label{sec:results}

\subsection{Performance and   Architectural Analysis}

As shown in Table~\ref{tab:results_50}, we first evaluate the models under the standard 50\% interference setting. Under this condition, the M-T-Hybrid achieves an SDR of 14.18 dB, comparing favorably to the VoViT baseline (13.76 dB). Notably, this performance is achieved with a substantially reduced parameter count of 16.2M compared to VoViT's 39.0M, suggesting that the hybrid Mamba-Transformer backbone offers an efficient architectural alternative to larger attention-based systems for audiovisual singing voice separation.
\begin{table}[t!]
\centering
\begin{tabular}{@{}lrrr@{}}
\toprule
\textbf{Model} & \textbf{Params (M)} & \textbf{SDR $\uparrow$} & \textbf{SIR $\uparrow$} \\
\midrule
\multicolumn{4}{@{}l}{\textit{Audio-Only}} \\
HT Demucs \cite{defossez2019demucs} & 18.9 & 6.07 & 10.07 \\
Audio-Only                          & 14.8 & 2.35 & 6.13 \\
\midrule
\multicolumn{4}{@{}l}{\textit{Audio-Visual}} \\
VoViT \cite{montesinos2022vovit}    & 39.0 & 13.76 & \textbf{21.84} \\
VoViT 100\%                         & 39.0 & 11.32 & 16.58 \\
Pure Mamba                          & 15.2 & 8.96  & 16.46 \\
Pure Transformer                    & 17.2 & 12.38 & 18.80 \\
T-M-Hybrid (Spec2Vec)               & 30.7 & 11.41 & 17.62 \\
M-T-Hybrid (Spec2Vec)               & 30.7 & 10.92 & 16.79 \\
T-M-Hybrid                          & 16.2 & 12.86 & 19.62 \\
M-T-Hybrid                          & \textbf{16.2} & \textbf{14.18} & 20.17 \\
\bottomrule
\end{tabular}
\caption{Separation performance and model complexity on Acapella's test-unseen dataset with \textbf{50\% interference}. M-T-Hybrid achieves comparable SDR with fewer parameters than VoViT. Results are in (dB).}
\label{tab:results_50}
\end{table}

To further investigate the robustness of the multimodal grounding mechanism, we conducted a stress test utilizing a 100\% interference ratio as seen in Table~\ref{tab:results_100}. This high-overlap scenario highlights a known limitation in audio-only paradigms: both the baseline and HT Demucs models experience a substantial drop in separation integrity, with SDR values near 0 dB. This behavior is typically linked to ``target swapping'', where the absence of a visual anchor makes it difficult to track a specific source during continuous spectral overlap. In this extreme setting, the M-T-Hybrid remains competitive; while it achieves slightly lower SDR than the reported VoViT baseline (11.11 dB vs. 11.82 dB), it maintains comparable separation under continuous interference. These results suggest that the visual modulation gate aids in preserving source identity when spectral cues alone are ambiguous.

\begin{table}[t!]
\centering
\begin{tabular}{@{}lcc@{}}
\toprule
\textbf{Model} & \textbf{SDR $\uparrow$} & \textbf{SIR $\uparrow$} \\
\midrule
\multicolumn{3}{@{}l}{\textit{Audio-Only}} \\
HT Demucs \cite{defossez2019demucs} & $-0.01$ & 4.92 \\
Audio-Only                          & 0.64    & 1.42 \\
\midrule
\multicolumn{3}{@{}l}{\textit{Audio-Visual}} \\
VoViT \cite{montesinos2022vovit}    & \textbf{11.82} & \textbf{19.37} \\
VoViT 100\%                         & 8.83    & 13.69 \\
Pure Mamba                          & 7.35    & 13.66 \\
Pure Transformer                    & 9.66    & 15.62 \\
T-M-Hybrid (Spec2Vec)               & 9.74    & 10.47 \\
M-T-Hybrid (Spec2Vec)               & 8.93    & 9.92 \\
T-M-Hybrid                          & 9.91    & 16.22 \\
M-T-Hybrid                          & 11.11   & 17.14 \\
\bottomrule
\end{tabular}

\caption{Separation performance on Acapella's test-unseen dataset with \textbf{100\% interference}. Results are in (dB) and best results are in bold ($N=1500$).}
\label{tab:results_100}
\end{table}

To validate the attention-based Band-Split encoder, we 
benchmarked it against a Spec2Vec configuration using uniform convolutional filters. As shown in Table 1, the M-T-Hybrid (Spec2Vec) (10.92 dB SDR) was significantly outperformed by the Band-Split version (14.18 dB). This confirms that musically-informed sub-band partitioning is superior to uniform filtering for capturing the harmonic fine-structure of singing voices.  As seen in Table 2, VoViT trained at 100 percent interference suffered a substantial drop to 8.83 dB SDR, compared to the standard VoViT's 11.82 dB. In contrast, our M-T-Hybrid maintained higher stability at 11.11 dB. 

The generalization of the architecture is further assessed using the URSing dataset (Table~\ref{tab:ursing_results}), which introduces different recording environments and dense backing vocals. The M-T-Hybrid demonstrates stability in this cross-domain setting, achieving a Mean SDR of 5.58 dB and a Median SDR of 10.18 dB. While the VoViT baseline shows a performance decrease on this dataset (1.83 dB Mean SDR), the hybrid configuration remains more consistent. The sustained SIR values suggest that the Mamba-Transformer sequence is capable of extracting generalizable vocal features, allowing the model to suppress cross-talk even when encountering significant domain shifts in the acoustic data.

\begin{table}[t!]
\centering
\scalebox{0.9}{
\begin{tabular}{lccc}
\hline
\textbf{Model} & \textbf{Mean SDR $\uparrow$} & \textbf{Median SDR $\uparrow$} & \textbf{SIR $\uparrow$} \\ \hline
HT Demucs \cite{defossez2019demucs} & 4.39 & 6.49 & 6.79 \\ 
Audio Only & 0.02 & 2.52 & 5.02 \\
\hline
VoViT \cite{montesinos2022vovit} & 1.83 & 4.64 & 2.57 \\ 
Pure Mamba & -0.09 & 2.98 & 8.75 \\
Pure Transformer & 3.86 & 7.58 & 11.10 \\
T-M-Hybrid & 5.07 & 9.25 & \textbf{11.53} \\
M-T-Hybrid & \textbf{5.58} & \textbf{10.18} & 11.46 \\ \hline
\end{tabular}
}

\caption{Separation performance on the URSing dataset with \textbf{50\% interference} (random extra vocal mixing, $N=959$). Best results are in bold. Results are in (dB).}
\label{tab:ursing_results}
\end{table}

\subsection{Computational Efficiency}

\begin{table}[t]
\centering
\begin{tabular}{@{}lcc@{}}
\toprule
\textbf{Model} & \textbf{Network inference time (ms)} \\ \midrule
VoViT \cite{montesinos2022vovit}          & 11.57  \\
Pure Transformer  & 5.01  \\
Pure Mamba        & 4.22  \\
M-T-Hybrid (Spec2Vec) & 9.92 \\
T-M-Hybrid (Spec2Vec) & 10.00 \\
T-M-Hybrid        & 3.95 \\
M-T-Hybrid        & \textbf{3.93} \\ 
\bottomrule
\end{tabular}
\caption{Separation network forward-pass time (ms) comparison. Measurements are averaged over 1500 samples using a 4-second audio input.}
\label{tab:latency}
\end{table}

Beyond separation metrics, we evaluate the computational efficiency of the model in Table~\ref{tab:latency}. These measurements report the forward-pass time of the separation network alone, and exclude facial landmark extraction, which runs as a separate AlphaPose pre-processing stage. Since VoViT relies on the same landmark pre-processing, excluding this stage affects both systems equally and the comparison remains like-for-like. The M-T-Hybrid achieves the lowest separation-network forward-pass time of 3.93 ms, representing a 2.94$\times$ speedup over VoViT (11.57 ms). Notably, the hybrid configuration is faster than both the Pure Transformer (5.01 ms) and Pure Mamba (4.22 ms) variants. This performance gain indicates that employing Mamba for initial long-range modeling efficiently compresses the sequence before Transformer-based refinement, mitigating the recurrent overhead associated with deep SSMs and the quadratic complexity of full-depth attention. These timings should therefore be interpreted as network inference time rather than full end-to-end system latency. A deployed system would also depend on the time required to accumulate an input chunk and to extract or track facial landmarks. Evaluating shorter chunks and robustness to unreliable face tracking remains an important direction for future work.

\subsection{Subjective Evaluation}

We conducted a listening test with 15 participants experienced in Music Source Separation, who rated Vocal Quality (Distortion) and Vocal Isolation (Interference suppression) on a 1–5 scale across 5 randomised 4-second excerpts from the test set.
Results (Table ~\ref{tab:mos_results}) support our objective metrics: the audio-only model and HT Demucs show low isolation scores (1.30, 1.83) due to target-swapping, while M-T-Hybrid achieves 3.88, comparable to VoViT (3.82). M-T-Hybrid also matches VoViT on quality (3.56 vs. 3.53), suggesting that combining Mamba's selective state-spaces with our Band-Split module offers an efficient alternative to heavier convolution-based architectures without sacrificing perceptual quality.
This test is preliminary. With only 15 participants and 5 excerpts each, the sample size cannot resolve the small gap between M-T-Hybrid and VoViT; their confidence intervals overlap. We therefore treat the two models as comparable, not one improved over the other. The gap between audio-visual and audio-only systems is the more robust finding. Confirming the AV comparison would need a larger study with formal significance testing.

\begin{table}[t]
\centering
\begin{tabular}{@{}lrr@{}}
\toprule
\textbf{Model} & \textbf{Vocal Quality} & \textbf{Vocal Isolation} \\
\midrule
HT Demucs  \cite{defossez2019demucs}              & 2.25 & 1.3 \\
Audio-Only           & 2.69 & 1.83 \\
VoViT \cite{montesinos2022vovit}                 & 3.53 & 3.82 \\
M-T-Hybrid    & \textbf{3.56} & \textbf{3.88} \\
\bottomrule
\end{tabular}
\caption{Subjective MOS results for Vocal Quality and Isolation. Min=1 / Max=5, the higher the better.}
\label{tab:mos_results}
\end{table}
\section{Conclusion}
\label{sec:conclusion}



We introduced MambaVoice, a novel hybrid Mamba-Transformer architecture designed for lightweight audiovisual singing voice separation. By utilizing a band-split audio encoder and integrating Mamba’s selective state-spaces with Transformer-based spectral refinement, our model overcomes the quadratic computational bottleneck of pure attention systems while maintaining the long-range temporal tracking necessary for complex vocal contours and dense spectral overlap. MambaVoice achieved competitive performance on the Acappella benchmark with 14.18 dB SDR, while utilizing 58\% fewer parameters than leading transformer baselines. Furthermore, the model's robust performance on the URSing dataset confirms its ability to generalize across diverse recording environments  without additional domain-specific tuning. MambaVoice achieves the lowest separation-network inference time among the evaluated configurations, at 3.93 ms. Future work will explore longer-context training, deeper cross-modal fusion strategies and extend it to other musical traditions and audio-visual speech separation.

\section{Acknowledgements}
\label{sec:acknowledgment}

This work is supported by the "Cátedra IA y Música" project (TSI-100929-2023-1), funded by the Secretaría de Estado de Digitalización e Inteligencia Artificial, the European Union-Next Generation EU funds and BMAT Music Innovators and by the "IMPA" project (PID2023-152250OB-I00) funded by MCIU/AEI/10.13039/501100011033/FEDER, UE. We would also like to thank the 15 participants who undertook the perceptual test.

\bibliography{ISMIRtemplate}

\end{document}